\documentclass[prb, twocolumn, amssymb, amsmath,aps,superscriptaddress,showpacs]{revtex4}
\usepackage{natbib}
\usepackage{colordvi}
\usepackage[dvips]{color}
\usepackage{graphicx}
\usepackage{epstopdf}
\usepackage{bm}
\usepackage{amsmath, amsfonts}
\usepackage{hyperref}
\begin{document}

\title{Strongly renormalized quasi-two-dimensional electron gas in a heterostructure 
with correlation effects}
\author{Andreas R\"uegg}
\affiliation{Theoretische Physik, ETH Z\"urich, CH-8093 Z\"urich}
\author{Sebastian Pilgram}
\affiliation{Theoretische Physik, ETH Z\"urich, CH-8093 Z\"urich}
\author{Manfred Sigrist}
\affiliation{Theoretische Physik, ETH Z\"urich, CH-8093 Z\"urich}
\affiliation{Department of Physics, Kyoto University, Kyoto 606-8502, Japan}

\date{\today}                                           

\begin{abstract}
We present aspects of strong electron correlation in a model of a BI/M/BI heterostructure where BI is a band insulator and M can be tuned from a metal to a Mott insulator by varying the on-site repulsion. An effective one-dimensional Schr\"odinger equation for the low-energy part is derived in the framework of the Kotliar-Ruckenstein slave-boson mean-field theory and solved self-consistently for a wide range of the on-site interaction strength and of the width of the sandwiched material M. In the strongly interacting limit coherent quasiparticles responsible for metallic behavior are confined to a relatively narrow region near the interfaces where they form a quasi-two-dimensional electron gas. Due to the lack of spatial homogeneity these quasiparticles intrinsically  obtain a nontrivial dependence on the transverse momentum which for example enters the low-energy behavior of the optical conductivity.
\end{abstract}
\maketitle

\section{Introduction}
Strongly correlated electron systems are strong candidates for the next generation functional materials for various purposes. This bright perspective is motivated by the richness of their phase diagrams involving a variety of ordered phases in charge, spin and orbital degrees of freedom, metal to insulator transitions, unconventional and high-$T_c$ superconductivity as well as various exotic liquid phases.\cite{Imada:1998qy} The electronic and magnetic phases show a high sensitivity to external conditions such as temperature, pressure, doping and application of electric or magnetic fields. In artificially built heterostructures involving different strongly correlated electron systems the conditions may change locally leading to new physics at the interface as compared to  the bulk material. These systems are likely to have remarkable properties which may be tuned by the application of electric fields\cite{thiel:2006a} or by modulation doping near the interfaces.\cite{lee:075106} Taking advantage of new mechanisms at interfaces of correlated electron systems may allow to design junctions for potential device applications, e.g., a colossal electroresistance device.\cite{oka:266403}

One class of recently investigated heterostructures is made of complex oxides which are insulating in the bulk systems.\cite{Ohtomo:2002fk, Ohtomo:2004lr, Maekawa:2006} In some cases, when brought together, a metallic quasi-two-dimensional electron gas at the interface is formed. Pioneering experimental work by Ohtomo and coworkers are performed at LaTiO$_3$/SrTiO$_3$ (Ref.~\onlinecite{Ohtomo:2002fk}) and at LaAlO$_3$/ SrTiO$_3$ (Ref.~\onlinecite{Ohtomo:2004lr}) interfaces and superlattices. In both cases the bulk materials are insulating but in the epitaxial heterostructures the transport properties can change dramatically. Thiel \emph{et al}.\cite{thiel:2006a} reported on a very large electric field response of the quasi-two-dimensional electron gas at the LaAlO$_3$/SrTiO$_3$ interface which is of potential interest for device applications. 

Our work is mainly inspired by the experiments on the  LaTiO$_3$/SrTiO$_3$ interface.\cite{Ohtomo:2002fk, Shibuya:2004} As pointed out by Okamoto and Millis\cite{Okamoto:2004uq} the atomically precise fabrication, the near lattice match ($a\approx 3.9$~\AA) and the chemical similarity of the two components offer a good starting point for theoretical studies and make it possible to study the influence of \emph{electronic reconstruction} alone:  how does the electronic phase at the interface differ from that in the bulk system? The pure La-compound is a Mott insulator with a Ti-$3d^1$ configuration whereas the pure Sr-compound is a band insulator (Ti-$3d^0$ configuration). When brought into contact the mutual doping of band and Mott insulator leads to a quasi-two-dimensional electron gas confined in an approximately $2$ nm thick region at the interface.\cite{Ohtomo:2002fk} Mainly for single-La-layer superlattices, where the effect of strong electron correlation is believed to be minor, density-functional theory calculations were successfully performed \cite{okamoto:056802, popovic:176805, hamann:195403, pentcheva:2006a} to primarily study lattice-relaxation effects and correlation-driven spin and charge ordering. In order to include additional effects of electron correlation Okamoto and Millis proposed a generalized Hubbard model for this system and combined insights from the Hartree approximation\cite{Okamoto:2004uq, okamoto:075101} and dynamical mean-field theory (DMFT).\cite{okamoto:241104, okamoto:235108} Focus was put on the electronic charge profile which was determined in the experiment by electron energy loss measurements and on possible ferromagnetic and antiferromagnetic ordering.  
 
The above discussed systems built an oxide analogy to more conventional semiconductor heterostructures. But strong on-site interaction in combination with a spatially nonuniform setup is expected to lead to differences as compared to both the strongly correlated bulk systems or heterostructures where electron correlations play a minor role. In order to give a systematic discussion on the influence of the on-site interaction we study a simplified model for heterostructures which are characterized by a BI/M/BI stacking. Here BI is a band-insulator such as SrTiO$_3$. By varying the on-site repulsion we can continuously change the properties of the sandwiched material M tuning it from a metal to a Mott insulator such as LaTiO$_3$. The model is based on a generalized single-band Hubbard model introduced in Ref.~\onlinecite{okamoto:241104}. So far, it was studied within the Hartree,\cite{okamoto:241104} Thomas-Fermi\cite{lee:075106}, and DMFT approximation.\cite{okamoto:241104,lee:075106} While the Hartree and Thomas-Fermi approximation are expected to give reasonable results for the charge distribution, i.e.\ the self-consistent screening, these approaches can not take into account the renormalization of the quasiparticle weight and some aspects of correlation are thus treated poorly. DMFT on the other hand is a powerful tool to describe these aspects but especially for spatially nonuniform systems a full treatment becomes numerically very involved. Here we report on an alternative approach to investigate the quasiparticles in such systems, namely, by means of the slave-boson mean-field (SBMF) approximation. The Kotliar-Ruckenstein (KR) slave-boson theory was originally formulated for the single-band Hubbard model.\cite{Kotliar:1986kx} It reproduces the Gutzwiller approximation\cite{Vollhardt:1984fk} on the mean-field level and is thus able to catch important aspects of strong correlation such as the suppression of the quasiparticle weight and at half filling the Brinkmann-Rice transition\cite{Brinkman:1970lr} from a paramagnetic metal to a paramagnetic insulator. Generalizations to a manifest spin-rotation invariant slave-boson representation were made by Li \emph{et al}.\cite{Li:1989lr} and to a unified spin- and charge-rotation invariant formulation by Fr\'esard and W\"olfle\cite{Fresard:1992a} leading to a more profound understanding of the KR slave-boson theory.  Applications to the orbital degenerated Hubbard model were performed to study (anti)ferromagnetism \cite{Fresard:1997uq, Hasegawa:1997fj} and the orbital-selective Mott transition.\cite{rueegg:2005}

The paper is organized as follows. In Sec.~\ref{sec:model} we introduce the model and apply the SBMF approximation. In Sec.~\ref{sec:res} we discuss the $T=0$ properties of the paramagnetic phase with focus on the coherent quasiparticles. Due to the lack of spatial homogeneity the quasiparticles obtain a non trivial momentum dependence which is studied by calculating the renormalization of the quasiparticle dispersion and the optical conductivity. In the last section a summary and conclusions are given.

\section{Model and Formalism}\label{sec:model}

\subsection{Model of a BI/M/BI heterostructure}
\begin{figure}
\centering
\includegraphics[width=0.8\linewidth]{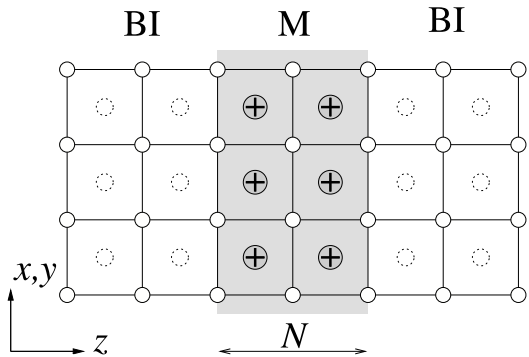}
\caption{Sketch of the heterostructure studied in this paper. BI is a band insulator and M is a material which can be tuned from a paramagnetic metal to a paramagnetic Mott insulator by varying the on-site repulsion. The heterostructure is defined by $N$ positively charged counter ions sitting in between the electronic sites.}\label{fig:model}
\end{figure}

We study the model of a [001] BI/M/BI heterostructure where BI is a band insulator and M a material which can be tuned from a metal to a Mott insulator by varying the on-site repulsion. As compared to the experimentally realized LaTiO$_3$/SrTiO$_3$ heterostructures the basic simplifications of the model are the disregard of the orbital degeneracy and the assumption of an unrelaxed perfect lattice match. A sketch of the studied system is shown in Fig.~\ref{fig:model}.
The heterostructure is defined by the electrostatic potential of $N$ ion-layers with simple positively charged ions sitting in between the electronic sites of a simple cubic lattice 
\begin{equation*}
V_i=-E_{\mathrm{C}}\sum_{j}\frac{1}{|\mathbf{r}_i-\mathbf{r}_{j}^{\mathrm{ion}}|}.
\end{equation*}
These counter ions simulate the electrostatic difference between Sr$^{2+}$ and La$^{3+}$. 
The long-range Coulomb interaction is characterized by the dielectric constant $\varepsilon_{\mathrm{D}}$, i.e., the Coulomb energy $E_{\mathrm{C}}=e^2/\varepsilon_{\mathrm{D}}a$. 
The model is described in terms of a generalized Hubbard Hamiltonian
\begin{equation}
\hat{H}=\hat{H}_{\mathrm{H}}+\hat{H}_{\mathrm{ion}} +\hat{H}_{\mathrm{ee}}
\label{eq:mod}
\end{equation}
with the ordinary Hubbard Hamiltonian including nearest-neighbor hopping and the on-site repulsion
\begin{equation*}
\hat{H}_{\mathrm{H}}=-t\sum_{\langle ij\rangle\sigma}\hat{c}_{i\sigma}^{\dag}\hat{c}_{j\sigma}^{\phantom\dag}+U\sum_i\hat{n}_{i\uparrow}\hat{n}_{i\downarrow}
\end{equation*}
and a part taking account of the long-range electron-ion and electron-electron Coulomb interaction
\begin{equation*}
\hat{H}_{\mathrm{ion}}+\hat{H}_{\mathrm{ee}}=\sum_iV_i\hat{n}_i+\frac{1}{2}\sum_{i\neq j}\hat{n}_iW_{ij}\hat{n}_j,
\end{equation*}
with the interaction matrix $W_{ij}=E_{\mathrm{C}}/{|\mathbf{r}_i-\mathbf{r}_j|}$.
For charge neutral systems, the model considered here [Eq.~(\ref{eq:mod})] involves three different parameters, namely $E_{\mathrm{C}}/t$, $N$ and $U/t$. The relative strength of the long-range Coulomb interaction $E_{\mathrm{C}}/t$ mainly affects the self-consistent screening of the ions. A realistic value for the experimental system realized by Othomo \emph{et al}.\cite{Ohtomo:2002fk} is $E_{\mathrm{C}}=0.8t$ corresponding to $\varepsilon_{\mathrm{D}}\approx15$, $a\approx3.9$~\AA \  and $t\approx0.3$~eV.\cite{Okamoto:2004uq, okamoto:075101} For the following discussion we fix $E_{\mathrm{C}}$ at this value. The relative on-site repulsion $U/t$ is intrinsically large for Mott insulators. Nevertheless we will treat this parameter continuously and consider the case where magnetic order is suppressed - therefore tuning the central material from a paramagnetic metal to a paramagnetic Mott insulator. Furthermore, we focus in our work on a systematic study on how the $T=0$  properties change by varying the width $N$ of the Mott-insulating material.

\subsection{Mean-field treatment}
One way to arrive at an effective low-energy theory is provided by the KR SBMF approximation. This can be done in full analogy to the homogeneous single-band Hubbard model.\cite {Kotliar:1986kx} We thus reformulate the model Hamiltonian [Eq.~(\ref{eq:mod})] in an enlarged Hilbert space by introducing  pseudo fermionic operators $\hat{f}_{i\sigma}^{(\dag)}$ and auxiliary bosonic operators (slave bosons) $\hat{e}_i^{(\dag)}$, $\hat{p}_{i\sigma}^{(\dag)}$ and $\hat{d}_i^{(\dag)}$ standing for the creation (annihilation) of empty, singly occupied sites with spin $\sigma$ and doubly occupied sites, respectively. Yet, this new Hilbert space is much larger than the Hilbert space of (\ref{eq:mod}). To project on the physical subspace we need to fulfill the following constraints for each individual site
\begin{eqnarray}
\hat{e}_i^{\dag}\hat{e}_i^{\phantom\dag}+\sum_{\sigma}\hat{p}_{i\sigma}^{\dag}\hat{p}_{i\sigma}^{\phantom\dag}+\hat{d}_i^{\dag}\hat{d}_i^{\phantom\dag}&\equiv&\hat{1}, \label{eq:conI} \\
\hat{p}_{i\sigma}^{\dag}\hat{p}_{i\sigma}^{\phantom\dag}+\hat{d}_i^{\dag}\hat{d}_i^{\phantom\dag}&\equiv&\hat{f}_{i\sigma}^{\dag}\hat{f}_{i\sigma}^{\phantom\dag}.
\label{eq:conQ}
\end{eqnarray}
In the mean-field approximation the slave-boson fields are replaced by their classical values and the constraints [Eq.~(\ref{eq:conI},\ref{eq:conQ})] are fulfilled on the average. We will concentrate  in the following on the paramagnetic mean-field solution. This allows to discuss important aspects of the Mott physics inherent in the model. In what follows we will use the convention  $i=(j,l)$ where $l$ is the layer index ($z$-direction) and $j$ specifies the site in each individual layer ($x,y$-direction). We fix the origin $z=0$ in the center of the Mott-insulating material and set the lattice spacing $a=1$. As independent mean fields we choose the amplitude of doubly-occupied sites $d_l$ and the total charge distribution $n_l$ in layer $l$ as well as a Lagrange multiplier $\lambda_l$ which controls the self-consistency of the charge distribution. The mean-field Hamiltonian can be diagonalized by a canonical transformation 
\begin{equation*}
\hat{f}_{jl\sigma}=\frac{1}{\sqrt{N_{||}}}\sum_{k\nu}\psi_{k\nu}(l)e^{ikR_{j}}\hat{f}_{k\nu\sigma}
\end{equation*}
which is composed of an in-plane Fourier transformation and of a transformation to the solutions of an effective one-dimensional Schr\"odinger equation in the $z$-direction
\begin{eqnarray}
\Big[(z_l^2-1)\varepsilon_k+\lambda_l\Big]\psi_{k\nu}(l)&-& t \sum_{\gamma=\pm1}z_{l}z_{l+\gamma}\psi_{k\nu}(l+\gamma) \nonumber \\
  &=&\left(E_{\nu k}-\varepsilon_k\right)\psi_{k\nu}(l).
\label{eq:Heff}
\end{eqnarray}
The quantum number $\nu=1,2,\dots$ is the band index, i.\ e.\ it labels the solutions of Eq.~(\ref{eq:Heff}) for fixed $k$. The bound states of Eq.~(\ref{eq:Heff}) define bands of a quasi-two-dimensional metal. 
Furthermore, the symmetry $z\leftrightarrow -z$ allows to choose eigenstates with a fixed parity $(-1)^{\nu-1}$,
\begin{equation}
\psi_{k\nu}(-l)=(-1)^{\nu-1}\psi_{k\nu}(l).
\label{eq:parity}
\end{equation}
The in-plane hopping is renormalized by a factor $z_l^2$ - corresponding to a layer dependent mass-renormalization $m_l^*/m=1/z_l^2$ -  and the hopping between different layers by $z_lz_{l'}$ where \cite {Kotliar:1986kx}
\begin{equation}
z_l(n_l,d_l)=\frac{\sqrt{(1-n_l+d_l^2)(n_l-2d_l^2)}+d_l\sqrt{n_l-2d_l^2}}{\sqrt{n_l(1-n_l/2)}}.
\label{eq:zl}
\end{equation}
In Eq.~(\ref{eq:Heff}), the inhomogeneous setup leads to an effective one-dimensional potential $(z_l^2-1)\varepsilon_k+\lambda_l$ along the heterostructure which can vary essentially for different values of the transverse momentum $k$, i.e., the in-plane kinetic energy $\varepsilon\equiv\varepsilon_k=-2t(\cos k_x+\cos k_y)$. The diagonalization of Eq.~(\ref{eq:Heff}) therefore has to be performed for each $\varepsilon_k$ separately.

The mean fields $d_l$, $n_l$ and $\lambda_l$ are determined by the stationary point of the free energy density at the inverse temperature $\beta=1/T$,
\begin{eqnarray}
& &f(d,n,\lambda)=-\frac{2}{\beta N_{||}}\sum_{k\nu}\ln(1+e^{-\beta E_{k\nu}})\label{eq:free}\\
& &+U_{\mathrm{r}}\sum_ld_l^2+\frac{1}{2}\sum_{ll'}n_lW_{ll'}n_{l'}-\sum_l(\lambda_l-V_l) n_l,\nonumber
\end{eqnarray}
under the constraint of charge neutrality $\sum_l n_l=N$. The saddle-point equations incorporating this condition are nonlinear, coupled self-consistency equations for the mean-fields:
\begin{eqnarray}
0&=&\bar{n}_l -n_l, \label{eq:selfnP}\\
0&=&U_{\mathrm{r}}d_l+\frac{\partial z_{l}^2}{\partial d_l}\bar{\varepsilon}_{l}-t\sum_{\gamma=\pm1}z_{l+\gamma}\frac{\partial z_{l}}{\partial d_l}\bar{\chi}_{l}^{(\gamma)}, \label{eq:selfdP}\\
\mu&=&-\lambda_l+V_l+\sum_{l'}W_{ll'}n_{l'}+2\frac{\partial z_{l}^2}{\partial n_l}\bar{\varepsilon}_{l}\nonumber\\
& &-2t\sum_{\gamma=\pm1}z_{l+\gamma}\frac{\partial z_{l}}{\partial n_l}\bar{\chi}_{l}^{(\gamma)}\label{eq:selflamP},
\end{eqnarray}
where we have introduced the following definitions
\begin{eqnarray}
\bar{n}_l&=&\frac{2}{N_{||}}\sum_{k\nu}\psi_{k\nu}(l)^2f_T(E_{k\nu}),\\
\bar{\varepsilon}_l&=&\frac{1}{N_{||}}\sum_{k\nu}\varepsilon_k \psi_{k\nu}(l)^2f_T(E_{k\nu}), \\
\bar{\chi}_l^{(\gamma)}&=&\frac{2}{N_{||}}\sum_{k\nu}\psi_{k\nu}(l) \psi_{k\nu}(l+\gamma)f_T(E_{k\nu}),
\end{eqnarray}
with the Fermi function $f_T(\omega)=(1+e^{\omega/T})^{-1}$. The saddle-point equations (\ref{eq:selfnP})-(\ref{eq:selflamP}) together with the effective Schr\"odinger equation (\ref{eq:Heff}) are the basic equations in our approach.

We remark here that in the physical subspace the operator identity 
\begin{equation}
U\hat{n}_{i\uparrow}\hat{n}_{i\downarrow}=(U-U_0)\hat{d}^{\dag}_i\hat{d}^{\phantom\dag}_i+U_0\hat{n}_{i\uparrow}\hat{n}_{i\downarrow}
\label{eq:freedom}
\end{equation}
holds with an arbitrary constant $U_0$.  However, the mean-field results depend on the choice of  $U_0$. The freedom resulting from Eq.~(\ref{eq:freedom}) is not unique to the particular model considered here but for example also exists in the SBMF approximation of the homogeneous Hubbard model. However, in the latter case we demand that for $U=0$ the non-interacting case is reproduced by the SBMF approximation\cite{Kotliar:1986kx} and consequently $U_0=0$. In the model considered here we find it useful to choose a finite threshold $U_0=E_{\mathrm{C}}$. Because usually the physical relevant parameter regime is $U\geq E_{\mathrm{C}}$ we define a \emph{minimal interacting} limit by $U=E_{\mathrm{C}}$ and demand that in this limit the SBMF approximation reproduces the Hartree mean field approximation. Hence, we set $U_0=E_{\mathrm{C}}$ meaning that double occupancy is only significantly reduced for $U>E_{\mathrm{C}}$. In our approach the relevant parameter describing the correlation induced deviation from the Hartree mean-field results is therefore $U_{\mathrm{r}}=U-E_{\mathrm{C}}$.

Including the shift for the on-site term discussed above the long-range Coulomb interaction matrix in units of $E_{\mathrm{C}}$ is given by
\begin{eqnarray}
\frac{W_{ll'}}{E_{\mathrm{C}}}&=&\sum_{(nm)}\frac{(1-\delta_{ll'})}{\sqrt{n^2+m^2+(l-l')^2}} \nonumber\\
&&+\delta_{ll'}\left(\frac{1}{4}+\sum_{(nm)\neq(00)}\frac{1}{\sqrt{n^2+m^2}}\right).\label{eq:Cmat}
\end{eqnarray}
Since we consider only charge neutral systems the divergent part of Eq.~(\ref{eq:Cmat}) is canceled by the divergent part of the long-range ion-electron interaction.

\subsection{Numerical scheme}
One way to solve the nonlinear saddle-point equations is first maximizing the free energy $f(n,d,\lambda)$ with respect to $\lambda$ and then minimizing the resulting function with respect to $d$ and $n$ (see, e.g., Ref.~\onlinecite{Moller:1993a}). It turns out that this strategy leads to a stable algorithm in a wide range of parameters whereas an iterative solving of the saddle-point equations fails for larger values of $U$ due to the strong nonlinearity. Thus, our algorithm contains basically two loops where in the first loop the free energy is maximized with respect to $\lambda$ and in the second loop the resulting function is minimized with respect to $d$ and $n$. 
The calculation of the free energy and of its gradients in the thermodynamic limit $N_{||}\rightarrow\infty$ involves a two-dimensional $k$-integration which can be reduced to a one-dimensional energy-integration including the density of states of the two-dimensional nearest neighbor tight-binding model 
\begin{equation}
N(\varepsilon)=\left\{\begin{array}{clcl}
(1/2\pi^2t)K\left( 1-(\varepsilon/4t)^2 \right) & &\mbox{ if $|\varepsilon |\leq 4t$;}\\
0 & & \mbox{else.} \end{array} \right.
\label{eq:freeDOS}
\end{equation}
Here, $K(x)$ is the complete elliptic integral of the first kind. Note that there is a van-Hoove singularity at $\varepsilon=0$.
For the diagonalization of Eq.~(\ref{eq:Heff}) we impose open boundary conditions and take typically about 30 layers into account, depending on the number of ion layers $N$. All the calculations are performed at $T=0$. The fact that only the band energy $\varepsilon$ enters formally into the SBMF 
analysis means also that we can treat various lattice topologies in the exactly same framework with 
modified density of states. 

\section{Ground-state properties}\label{sec:res}

\subsection{Double occupancy and charge distribution}
\begin{figure}
\centering
\includegraphics[width=0.8\linewidth]{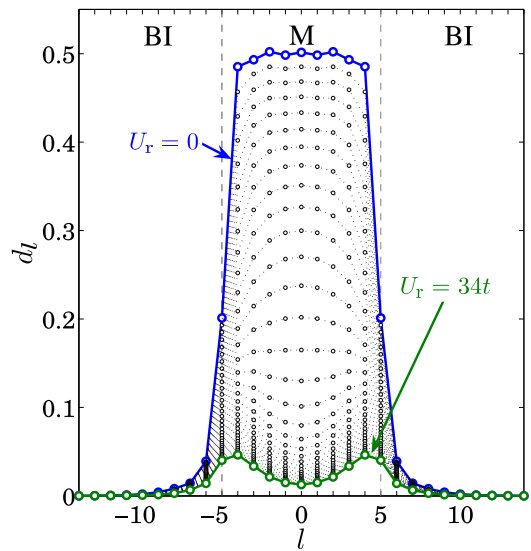}
\caption{(Color online) The mean-field $d_l$ for the $N=10$ ion-layers heterostructure for $U_{\mathrm{r}}=0,\dots,34t$. The fraction of doubly occupied sites in layer $l$ is given by $d_l^2$.}
\label{fig:dl}
\end{figure}
\begin{figure}
\centering
\includegraphics[width=0.9\linewidth]{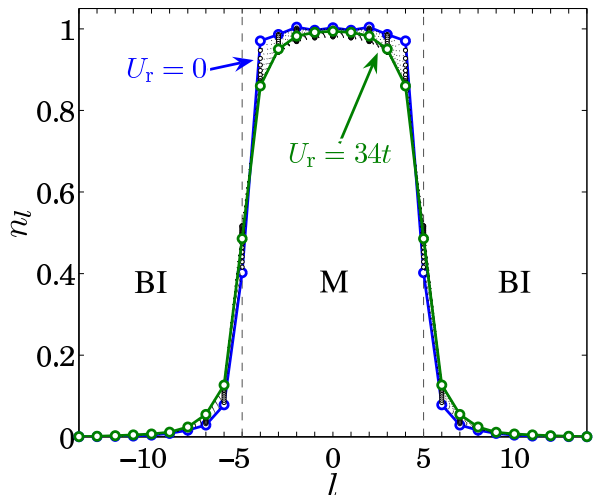}
\caption{(Color online) Charge distribution for the $N=10$ ion-layers heterostructure for various values of $U_{\mathrm{r}}$ between $0$ and $34t$.}\label{fig:chargedis}
\end{figure}

We first discuss the fraction of doubly occupied sites given by the mean field $d_l^2$ and the charge distribution $n_l$. It is clear, that in order to reduce the on-site potential energy cost, doubly occupied sites are suppressed by increasing $U_{\mathrm{r}}$. We show this in Fig.~\ref{fig:dl} for the $N=10$ heterostructure. In the minimal interacting case, $U_{\mathrm{r}}=0$, the fraction of doubly occupied sites in layer $l$ is $d_l^2=n_l^2/4$. With increasing $U_{\mathrm{r}}$, $d_l$ tends to zero. As expected, doubly occupied sites are strongest suppressed in the Mott-insulating material, namely in the center of the heterostructure where the charge is close to $1$ (see Fig.~\ref{fig:chargedis}). Consequently, the hopping renormalization factor $z_l$ [Eq.~(\ref{eq:zl})] is reduced there reflecting the suppression of charge fluctuations.  On the other hand, in the band-insulating material,  $d_l$ varies only little with increasing $U_{\mathrm{r}}$ and stays close to the minimal interacting value. Therefore, $z_l\approx 1$ in this region. 

In the paramagnetic phase discussed here, the system has two competing possibilities to reduce the double occupancy and thus the on-site energy cost:
\begin{itemize}
\item[(i)]{Spreading of the electron charge distribution.}
\item[(ii)]{Localization of the electronic states.}
\end{itemize}
Of course, in both cases, long-range Coulomb as well as kinetic energy has to be payed. Figure~\ref{fig:chargedis} shows the electron charge distribution $n_l$ of the $N=10$ heterostructure for different values of $U_{\mathrm{r}}$. We find a rather small dependence of $n_l$ on $U_{\mathrm{r}}$: the effect of the long-range Coulomb field dominates over the short-range correlations and consequently the above mentioned two mechanisms play a rather minor role for the overall growth direction electron density distribution.\cite{okamoto:075101,okamoto:241104,Okamoto:2004uq}
Nevertheless, a closer look at the evolution of the charge distribution allows to distinguish two different regimes characterized by one of the two mechanisms. Starting from the minimal interacting case and increasing $U_{\mathrm{r}}$ results in a spread of the particle density. This effect is also included in the Hartree approximation where the decoupling of the $U$-term leads to an additional term in the self-consistent electrostatic potential $\lambda_l\rightarrow \lambda_l+U_{\mathrm{r}}n_l/2$ which favors a homogeneous charge distribution. We therefore refer to this regime as the \emph{Hartree regime}.
\begin{figure}
\centering
\includegraphics[width=0.7\linewidth]{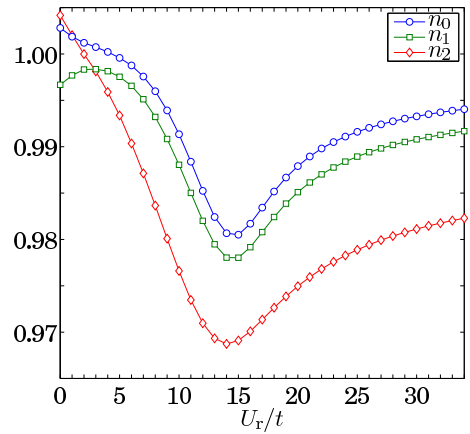}
\caption{(Color online) The occupancy of the central layers as a function of $U_{\mathrm{r}}$ for $N=10$.}
\label{fig:n0}
\end{figure}
But for all $N>1$ we observe that in most layers the density $n_l$ evolves non-monotonically  with increasing $U_{\mathrm{r}}$. Namely, there is a turning point where the slope of $n_l(U_{\mathrm{r}})$ changes sign. Figure~\ref{fig:n0} shows the occupancy of the central layers $l=0$, $l=\pm 1$ and $l=\pm 2$ of the $N=10$ heterostructure as a function of $U_{\mathrm{r}}$. The turning-point-value of $U_{\mathrm{r}}$ is different for the different layers and varies also with $N$ but is roughly in the range of the critical value $U_{\mathrm{c}}^{\mathrm{BR}}\approx 16t$ of the Brinkmann-Rice transition in the homogeneous system. We therefor refer to the regime $U_{\mathrm{r}}>U_{\mathrm{c}}^{\mathrm{BR}}$ as the \emph{Mott regime}.

\begin{figure}
\centering
\includegraphics[width=1\linewidth]{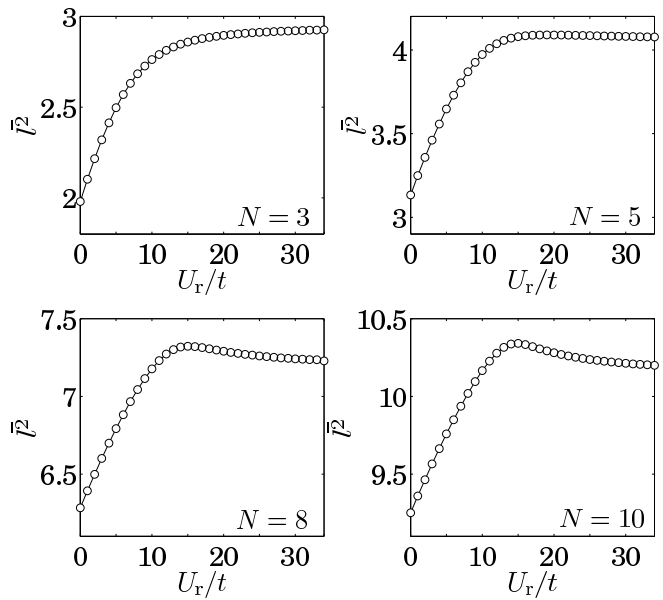}
\caption{The mean distance $\bar{l^2}$ of the charge from the center as defined in Eq.~(\ref{eq:meandis}) shown for $N=3,5,8,10$ as a function of $U_{\mathrm{r}}$.}
\label{fig:meandis}
\end{figure}
\begin{figure}
\centering
\includegraphics[width=0.8\linewidth]{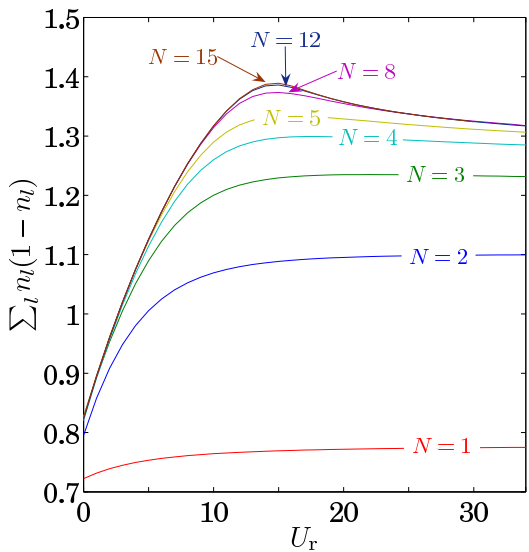}
\caption{(Color online) The quantity $\Delta=\sum_ln_l(1-n_l)$ measuring the width of the interface as function of $U_{\mathrm{r}}$ for different heterostructures $N$.}
\label{fig:width}
\end{figure}
These observations suggest to look at the following two quantities characterizing the charge distribution. We define the square mean distance of the charge from the center of the heterostructure as
\begin{equation}
\bar{l^2}(N,U_{\mathrm{r}})=\frac{\sum_ll^2n_l(N,U_{\mathrm{r}})}{N}
\label{eq:meandis}
\end{equation}
and a quantity measuring the width of the interface as 
\begin{equation}
\Delta=\sum_ln_l(1-n_l).
\end{equation}
$\Delta$ is a measure of the region where the electron density $n_l$ differs significantly from the bulk values $n_l=0$ (band insulator) and $n_l=1$ (Mott insulator). In the strongly interacting limit the mobile carriers are mainly confined to this region (see also Sec.~\ref{sec:coherent}).
Figure~\ref{fig:meandis} shows $\bar{l^2}$ for $N=3,5,8,10$ as a function of $U_{\mathrm{r}}$. In all cases we can clearly distinguish Hartree and Mott regime. In the Hartree regime we observe a rather steep increase of $\bar{l^2}$ reflecting that the reduction of the doubly occupied sites is mainly achieved by the spreading of the charge, namely by the mechanism (i). On the other hand, in the Mot regime,  $\bar{l^2}$ is approximatively constant (it flattens or even slightly decreases) indicating that in this regime the effect of localization is predominant [mechanism (ii)]. 
A similar behavior as function of $U_{\mathrm{r}}$ is also observed in the width $\Delta$ of the interface (see Fig.~\ref{fig:width}). In addition, for $N>5$, $\Delta$ becomes almost independent of $N$ and differs only in the vicinity of $U_{\mathrm{c}}^{\mathrm{BR}}$ reflecting that for large $N$ the two interfaces are disconnected. 

The behavior of $\bar{l^2}$ and $\Delta$ is a consequence of the competition between the three energy scales in the problem, namely $E_{\mathrm{C}}$, $U_{\mathrm{r}}$ and the bandwidth $W=12t$. In the Hartree regime $E_{\mathrm{C}}<U_{\mathrm{r}}\ll W$ the qualitative behavior observed in Fig.~\ref{fig:meandis} and \ref{fig:width} is basically determined by the competition between $E_{\mathrm{C}}$ and $U_{\mathrm{r}}$. The penetration depth thus increases by increasing $U_{\mathrm{r}}$ due to the  mechanism (i). On the other hand, in the Mott regime $E_{\mathrm{C}}<W\ll U_{\mathrm{r}}$ the penetration depth is mainly determined by the competition between $E_{\mathrm{C}}$ and $W$ and therefore essentially independent of $U_{\mathrm{r}}$. Because the electronic states are almost localized or, in other words, the binding of the electrons to the ions is strengthened, the charge distribution is almost independent of $U_{\mathrm{r}}$.

\subsection{Renormalization of the quasiparticle dispersion}
\begin{figure}
\centering
\includegraphics[width=1\linewidth]{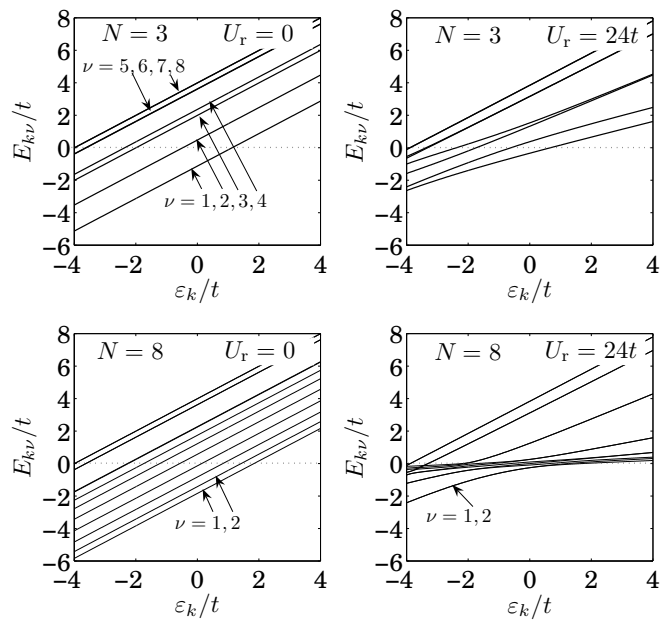}
\caption{The quasiparticle dispersion $E_{k\nu}$ of the partially filled bands in the $N=3$ and $N=8$ heterostructure as a function of $\varepsilon_k=-2t(\cos k_x+\cos k_y)$ both for $U_{\mathrm{r}}=0$ and $U_{\mathrm{r}}=24t$.}
\label{fig:dispN3}
\end{figure}
We now go over to a discussion of the solutions of the effective Schr\"odinger equation (\ref{eq:Heff}). Whenever it is appropriate, we will use the notation $E_{\nu}(\varepsilon)$ for the quasiparticle dispersion and $\psi_{\varepsilon\nu}(l)$ for the transverse part of the quasiparticle wave function because the momentum dependence enters only via $\varepsilon\equiv\varepsilon_k=-2t(\cos k_x+\cos k_y)$. In general we find $N+5$ partially filled bands in the ground state. In the minimal interacting limit ($U_{\mathrm{r}} =0 $), the dispersion of these bands has the form $E_{\nu}(\varepsilon)=E_{\nu}+\varepsilon$, i.e.\ $E_{\nu}(\varepsilon)$ has slope one. Increasing $U_{\mathrm{r}}$ leads to a renormalization of the quasiparticle dispersions and $E_{\nu}(\varepsilon)$ flattens, thus indicating localization of the electronic states. The slope of $E_{\nu}(\varepsilon_k)$ is given by
\begin{equation}
Z_{k\nu}=\frac{\partial E_{k\nu}}{\partial \varepsilon_k}=\sum_lz_l^2\psi_{k\nu}(l)^2
\label{eq:Zknu}
\end{equation}
which, as a consequence of the layer-dependent mass renormalization, is in general less than one.
 
As an example we consider here the $N=3$ heterostructure with $8$ partially filled bands and the $N=8$ heterostructure with 13 partially filled bands. The dispersion in the minimal interacting case and for $U_{\mathrm{r}}=24t$ is plotted in Fig.~\ref{fig:dispN3}. The particular geometry of the system, namely the symmetry $z\leftrightarrow -z$, allows that some of the quasiparticle bands are (almost) degenerate. We can distinguish two types of degeneracies: (1) Essentially independent of the interaction strength we find that the almost unbounded states (the highest partially filled bands) give rise to pairwise degenerate bands. In the case $N=3$ this is observed for the states $\nu=5,6$ and $\nu=7,8$.  A similar feature can be seen for the $N=8$ heterostructure where the six highest bands $\nu=8-13$ are pairwise degenerate. (2) In addition, by increasing $U_{\mathrm{r}}$, also the lowest bands can become pairwise degenerate in certain $k$-regions. This is observed in Fig.~\ref{fig:dispN3} for the $\nu=1,2$ and $\nu=3,4$ bands of the $N=8$ heterostructure and reflects that the two interfaces become disconnected from each other. The aspect of correlation induced degeneracy will be discussed in details in the next section.

The renormalization of the quasiparticle dispersion is, in general, the stronger the bigger $N$ and $U_{\mathrm{r}}$ and can substantially depend on $\nu$ and $\varepsilon = \varepsilon_k$. A qualitative understanding of the $k$-dependence is given by the following considerations. 
For the moment let us ignore the interlayer hopping and assume that the bulk system of the sandwiched material is Mott-insulating. Then the dispersion in each individual layer can be interpreted as an uniformly renormalized band $\epsilon_l(\varepsilon)=z_l^2\varepsilon +\lambda_l$.  Almost entirely flat bands near the Fermi energy, which are close to half filling, correspond to layers in the center of the heterostructure. On the other hand, more dispersive bands with low filling correspond to layers further away from the center. If we now allow for a finite interlayer hopping $\tilde{V}_{ll'}=-tz_lz_{l'}$ the uniformly renormalized bands hybridize. As a result, the renormalization of the hybridized bands can strongly depend on $\varepsilon= \varepsilon_k$. 
\begin{figure}
\centering
\includegraphics[width=1\linewidth]{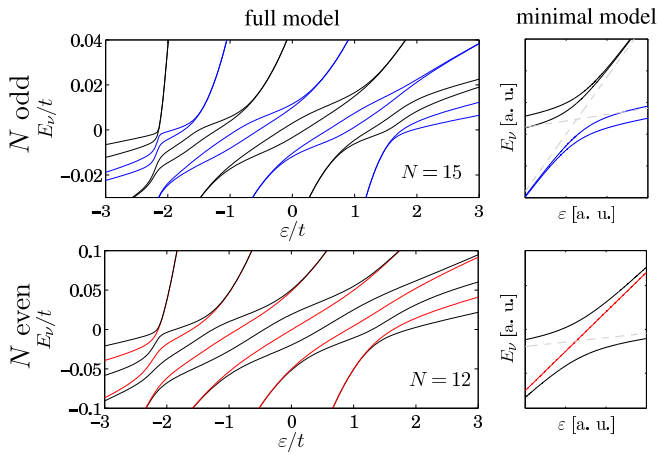}
\caption{(Color online) The quasiparticle dispersion $E_{k\nu}$ near the Fermi energy for $U_{\mathrm{r}}=24t$. Two types of avoided level crossing are observed depending on whether $N$ is even or odd. The type of avoided level crossing can be understood within a minimal model considering 3 ($N$ even) respectively 4 layers ($N$ odd).}
\label{fig:dispNALC}
\end{figure}

The $k$-dependency is strongest in the vicinity of the Fermi energy where $E_{\nu}(\varepsilon)$ shows  a complex pattern of avoided level crossings as function of $\varepsilon$ (see Fig.~\ref{fig:dispNALC}). Depending on whether $N$ is even or odd we can distinguish two types of avoided level crossing. The "building blocks" can be calculated within a minimal model including 3 ($N$ even) or 4 ($N$ odd) layers as shown in the Appendix~\ref{ap:minmod}.

The above discussion suggests a certain relation to the heavy-fermion ground state of the Periodic Anderson model.\cite{Tsunetsugu:1997fk} There, the hybridization between the dispersionless and strongly correlated $f$ level and the itinerant conductance band leads to a heavy Fermion liquid with strongly renormalized quasiparticle properties.\cite{Rice:1985lr} However, similar to the periodic Anderson model, in the model considered here the stability of the paramagnetic heavy-Fermi liquid state against magnetic ordering is an important question which will be addressed elsewhere.

Finally we remark here that the filling of the quasiparticle bands also changes by varying $U_{\mathrm{r}}$. However, it seems that this effect is rather weak and we refrain from discussing this point in more details.

\subsection{Correlation induced degeneracy and momentum dependence of $\psi_{k\nu}(l)^2$}

\begin{figure}
\centering
\includegraphics[width=0.8\linewidth]{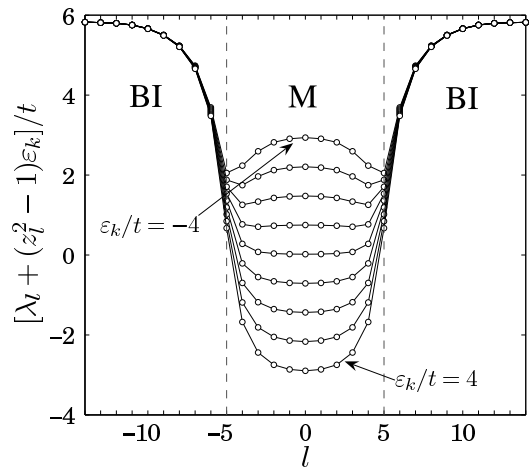}
\caption{The effective one-dimensional potential $\lambda_l+(z_l^2-1)\varepsilon_k$ for various values of $\varepsilon_k=-4t,-3t,\dots,4t$ for $U_{\mathrm{r}}=14t$ and $N=10$.}
\label{fig:effpot}
\end{figure}

\begin{figure}
\centering
\includegraphics[width=1\linewidth]{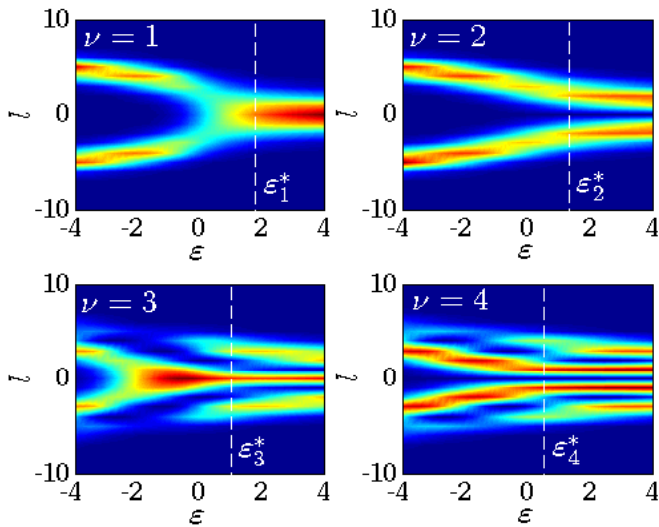}
\caption{(Color online) Density plot of the square of the four lowest lying transverse wave functions $\psi_{\varepsilon\nu}(l)^2$ as a function of $l$ and $\varepsilon$ for the same parameters as in Fig.~\ref{fig:effpot}. The Fermi energy is indicated as white dashed lines.}
\label{fig:mod}
\end{figure}

\begin{figure}
\centering
\includegraphics[width=1\linewidth]{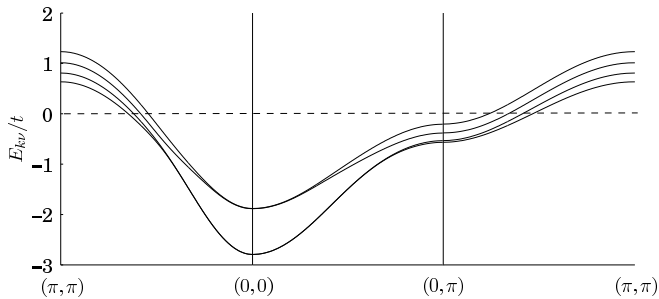}
\caption{The quasiparticle dispersion $E_{k\nu}$ of the four lowest lying bands in the $N=10$ heterostructure for $U_{\mathrm{r}}=14t$. Near $(k_x,k_y)\approx(0,0)$ they are pairwise degenerated.}
\label{fig:dispN10U14}
\end{figure}

Within the SBMF approximation the $k$-dependence of the quasiparticles formally enters through the term $(z_l^2-1)\varepsilon_k$ in the effective Schr\"odinger equation (\ref{eq:Heff}). This leads to a $\varepsilon$-dependent effective potential $\lambda_l+(z_l^2-1)\varepsilon$ which includes the screened ion-potential as well as a 'Mott potential' introducing an effective central barrier ($\to $ double-well potential form)  for $\varepsilon<0$ and a single well for $\varepsilon>0$.  Thus, for $k$-values in the center of the first Brillouin zone ($\varepsilon<0$), a double-well potential is formed and the regions $l>0$ and $l<0$ are almost decoupled. As an example we show $\lambda_l+(z_l^2-1)\varepsilon$ in Fig.~\ref{fig:effpot} for $N=10$ and $U_{\mathrm{r}}=14t$.
This means that electrons which have a low (in-plane) kinetic energy prefer to stay mainly in the interface region rather than in the center where the effect of the on-site interaction $U_{\mathrm{r}}$ is much stronger. To illustrate the rather complex dependence of the transverse part $\psi_{\varepsilon\nu}$ on $\varepsilon$, we show in Fig.~\ref{fig:mod} a density plot of $\psi_{\varepsilon\nu}(l)^2$ for $\nu=1, \dots, 4$ as a function of $\varepsilon$ and $l$ for the same parameters as in Fig.~\ref{fig:effpot}.  In this plot we interpolate $\psi_{\varepsilon\nu}(l)^2$ between different $l$'s. The $\varepsilon$ value of the Fermi energy is denoted by $\varepsilon_{\nu}^*$ and is indicated as a white dotted line. Due to the double-well structure the density plot of $\psi_{\varepsilon1}^2$ and $\psi_{\varepsilon2}^2$ looks very similar for $\varepsilon<0$.
As a consequence, energetically nearby states of different parity can become almost degenerated in a certain $\varepsilon$-range. Figure~\ref{fig:dispN10U14} shows the four energetically lowest lying bands of the $N=10$ heterostructure for $U_{\mathrm{r}}=14t$. 
These bands are pairwise degenerated near $(k_x,k_y)\approx(0,0)$. 

\subsection{Quasiparticle weight}
A characteristic quantity describing correlation effects is the quasiparticle weight $Z_{\nu}$ of the band $\nu$. It for examples enters the linear specific heat at low temperatures 
\begin{equation*}
c_v N=-T\frac{\partial^2 f}{\partial T^2}=\frac{\pi^2}{3}\sum_{\nu}\frac{2N(\varepsilon_{\nu}^*)}{Z_{\nu}}T,
\end{equation*}
where $E_{\nu}(\varepsilon_{\nu}^*)=0$ and $N(\varepsilon)$ is the free electron density of states [Eq.~(\ref{eq:freeDOS})]. The quasiparticle weight accounts for the renormalization of the Fermi velocity - a vanishing $Z_{\nu}$ thus indicates a localized state.  
It is given by the value of $Z_{k\nu}$ [Eq.~(\ref{eq:Zknu})] at the Fermi energy:
\begin{equation}
Z_{\nu}=\frac{\partial E_{\nu}}{\partial \varepsilon}(\varepsilon_{\nu}^*)=\sum_lz_l^2\psi_{\varepsilon_{\nu}^*\nu}(l)^2.
\label{eq:Znu}
\end{equation}
Here, $\psi_{\varepsilon_{\nu}^*\nu}$ is the transverse part of the wave function at the Fermi energy. The above expression will reappear later when we calculate the discontinuity in the momentum distribution function at the Fermi energy at $T=0$.

\begin{figure}
\centering
\includegraphics[width=0.8\linewidth]{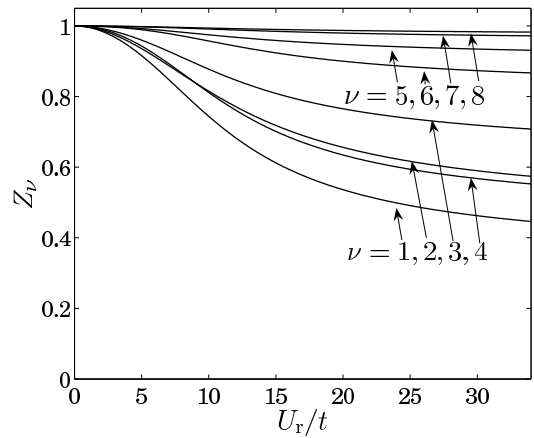}
\caption{Quasiparticle weight $Z_{\nu}$ of the partially filled bands $\nu=1,\dots ,8$ as a function of $U_{\mathrm{r}}$ for $N=3$.}
\label{fig:Znu}
\end{figure}

The quasiparticle weight $Z_{\nu}$ of the partially filled bands of the $N=3$ heterostructure is shown in Fig.~\ref{fig:Znu} as a function of $U_{\mathrm{r}}$. Energetically lower lying bands are in general more affected by $U_{\mathrm{r}}$ and the $\nu=1$ band has always the smallest quasiparticle weight. Note however that this band is \emph{not} closest to half filling. The exact order of $Z_{\nu}$ for $\nu>1$ depends on $N$ and in some cases also on $U_{\mathrm{r}}$. Because odd wave functions have nodes at $z=0$ a parity effect is visible. If $N$ is even the electronic sites are at $l=0,\pm1,\pm2,\dots$ and odd-parity bands are less renormalized because $\psi_{\varepsilon_{\nu}^*\nu}(0)=0$. On the other hand, if $N$ is odd, the electronic sites are at $l=\pm1/2,\pm3/2\dots$ and the odd-parity bands tend to be more strongly renormalized because the zero at $z=0$ lies between the electronic sites. For $N=3$, this parity effect is seen in Fig.~\ref{fig:Znu} where $Z_4<Z_3$ and $Z_6<Z_5$. We note here that both magnitude and order of the quasiparticle weight is in good qualitative agreement with the DMFT results shown in Fig.~3.(b) of Ref.~\onlinecite{okamoto:241104}.

\begin{figure}
\centering
\includegraphics[width=0.8\linewidth]{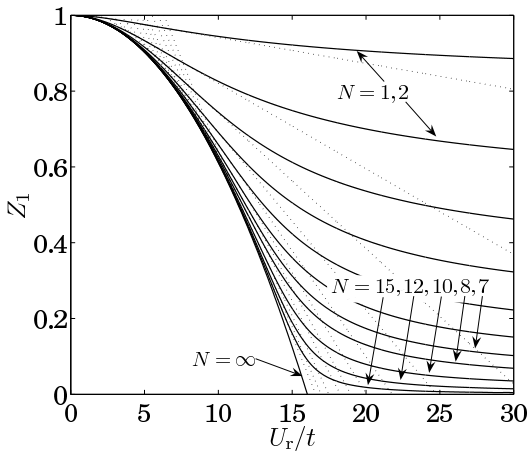}
\caption{The quasiparticle weight $Z_{1}$ for $N=1,2,\dots,8,10,12,15,\infty$ as a function of $U_{\mathrm{r}}$. In order to define a characteristic value $U_{\mathrm{r}}^{\mathrm{c}}(N)$ we extrapolate $Z_1$ at the inflection point to zero. }
\label{fig:Z1N}
\end{figure}

\begin{figure}
\includegraphics[width=0.7\linewidth]{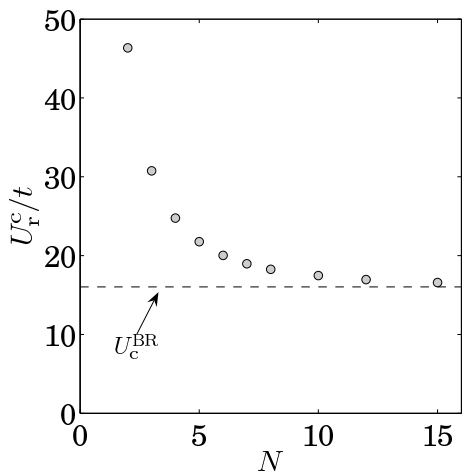}
\caption{$U_{\mathrm{r}}^{\mathrm{c}}$ as a function of $N$. The dashed line denotes the value of the Mott transition in the $N\rightarrow\infty$ model.}
\label{fig:Urc}
\end{figure}

As a further example we show in Fig.~\ref{fig:Z1N} the quasiparticle weight $Z_1$ of the lowest lying band as a function of $U_{\mathrm{r}}$ for various values of $N$. Note that for $N=1$ there is no significant renormalization of the quasiparticle weight which is in agreement with the observation made in density-functional theory based calculations for the $N=1$ heterostructure\cite{popovic:176805} and reflects the special role of this case\cite{Okamoto:075101}. In order to have a more quantitative description how on-site correlation effects increase with growing $N$ we define a characteristic value $U_{\mathrm{r}}^{\mathrm{c}}(N)$ for each $N$ by extrapolating $Z_1(U_{\mathrm{r}})$ at the inflection point to zero which yields $U_{\mathrm{r}}^{\mathrm{c}}(N)$. 
In Fig.~\ref{fig:Urc} we show $U_{\mathrm{r}}^{\mathrm{c}}(N)$ as a function of $N$. For $N\rightarrow\infty$ we find that $U_{\mathrm{r}}^{\mathrm{c}}$ saturates at $U_{\mathrm{c}}^{\mathrm{BR}}\approx 16t$, the critical interaction strength for the Mott transition in the bulk system of the sandwiched material.

\begin{figure}
\centering
\includegraphics[width=0.8\linewidth]{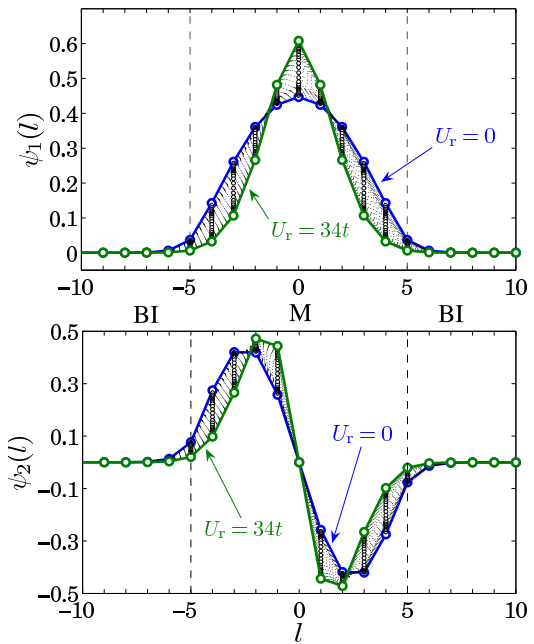}
\caption{(Color online) The transverse part of the wave function $\psi_{\varepsilon_{\nu}^*\nu}(l)$ at the Fermi energy for the two lowest lying bands for various values of $U_{\mathrm{r}}=0-34t$ for the $N=10$ heterostructure.}
\label{fig:Psis}
\end{figure}
Beside the suppression of $z_l$  in the center of the heterostructure there is also a slight change in $\psi_{\varepsilon_{\nu}^*\nu}$ with increasing $U_{\mathrm{r}}$. Namely, there is a clear tendency towards localization in the center of the heterostructure which reduces the quasiparticle weight $Z_{\nu}$ [Eq.~(\ref{eq:Znu})] even more. As an illustrative example, the wave functions at the Fermi energy of the $N=10$ heterostructure are shown in Fig.~\ref{fig:Psis} for $\nu=1,2$ and for a range of $U_{\mathrm{r}}=0,\dots,34t$.

Finally, we point out that we do not observe a strict Mott transition for finite $N$ and $U_{\mathrm{r}}$ in the sense of a vanishing quasiparticle weight $Z_{\nu}$ respectively of a divergent effective mass $m_l^*\sim1/z_l^2$. Instead, the mutual doping of Mott and band insulator leads to metallic behavior of the whole structure. The system gains energy by keeping a certain amount of charge fluctuation meaning that $Z_{\nu}>0$. In other words, the hopping between the individual layers is never renormalized to $0$ and $n_l<1$ and $z_l>0$. In the spatially nonuniform system discussed here, there is a large amount of freedom to reduce the on-site energy cost.  Among all the states characterized by $(k\nu)$ only a part is strongly affected by the correlations. This allows for an optimal reduction of the on-site energy cost without  cutting back much from the kinetic or long range Coulomb energy. Thus, the possibility that correlation effects can be selectively split among a large number of different states prevents the system from undergoing a strict Mott transition.

\subsection{Layer resolved spectral density and coherent particle density}
\label{sec:coherent}
The dynamical properties of the correlated electrons in the heterostructure can be best studied by looking at the single particle spectral function. In principle, the spectral functions are accessible in photoemission\cite{takizawa:057601} or scanning tunneling microscopy. In the framework of the SBMF approximation only the coherent (low-energy) part of the spectral weight can be captured. In order to access to lower and upper Hubbard bands within slave-boson theory one has to go beyond the saddle-point approximation\cite{Raimondi:1993lr} which is beyond the scope of this paper. Nonetheless, the low-energy part is important for the metallicity of the heterostructure. 

The $k$-dependent layer resolved spectral density is defined as the imaginary part of a retarded one-particle Greens function
\begin{equation}
A_{lk\sigma}(\omega)=-\frac{1}{\pi}\mathrm{Im}\, G_{llk\sigma}(\omega).
\end{equation}
Within the SBMF approach the \emph{coherent} part of $G_{llk\sigma}(\omega)$ is given by
\begin{equation}
G_{llk\sigma}^{\mathrm{coh}}(\omega)=\sum_{\nu}\frac{z_l^2\psi_{k\nu}(l)^2}{\omega-E_{k\nu}+i0^+}
\end{equation}
and we find for the coherent part of the spectral density
\begin{equation}
A_{lk\sigma}^{\mathrm{coh}}(\omega)=\sum_{\nu}z_l^2\psi_{k\nu}(l)^2\delta(\omega-E_{k\nu}).
\end{equation}
Integration of $A_{lk\sigma}(\omega)f_0(\omega)$ over $l,\omega$ yields the ground-state momentum distribution function of the correlated electrons. The coherent part of the momentum distribution function $n_{k\sigma}^{\mathrm{coh}}=\sum_{\nu}n_{k\nu\sigma}^{\mathrm{coh}}$ contains contributions of the different bands
\begin{equation}
n_{k\nu\sigma}^{\mathrm{coh}}\equiv n_{k\nu}^{\mathrm{coh}}=Z_{k\nu}f_0(E_{k\nu})
\label{eq:nkcoh}
\end{equation}
where $Z_{k\nu}$ is given in Eq.~(\ref{eq:Zknu}). This means that each state $(k\nu\sigma)$ is weighted by a factor $Z_{k\nu}\leq 1$. Therefore, also the discontinuity of $n_{k\nu\sigma}^{\mathrm{coh}}$ at the Fermi energy is reduced, namely, it is equal to $Z_{\nu}=\sum_lz_l^2\psi_{\varepsilon_{\nu}^*\nu}(l)^2=\partial E_{\nu}(\varepsilon_{\nu}^*)/\partial \varepsilon$ and we recover the expression for the quasiparticle weight given in Eq.~(\ref{eq:Znu}).

The ($k,\sigma$ integrated) layer resolved spectral density can be expressed in terms of the free electron density of states $N(\varepsilon)$ [Eq.~(\ref{eq:freeDOS})]  and is given by
\begin{eqnarray}
A_l^{\mathrm{coh}}(\omega)&=& \frac{1}{N_{||}}\sum_{k\sigma}A_{lk\sigma}^{\mathrm{coh}}(\omega) \nonumber \\
&=& 2z_l^2\sum_{\nu}\frac{N(\varepsilon_{\nu\omega})\psi_{\varepsilon_{\nu\omega}\nu}(l)^2}{Z_{\nu\omega}}.
\label{eq:Alomega}
\end{eqnarray}
Here, $\varepsilon_{\nu\omega}$ is defined by the equation $E_{\nu}(\varepsilon_{\nu\omega})=\omega$ and $Z_{\nu\omega}=\partial E_{\nu}(\varepsilon_{\nu\omega})/\partial \varepsilon$.

\begin{figure}
\centering
\includegraphics[width=1\linewidth]{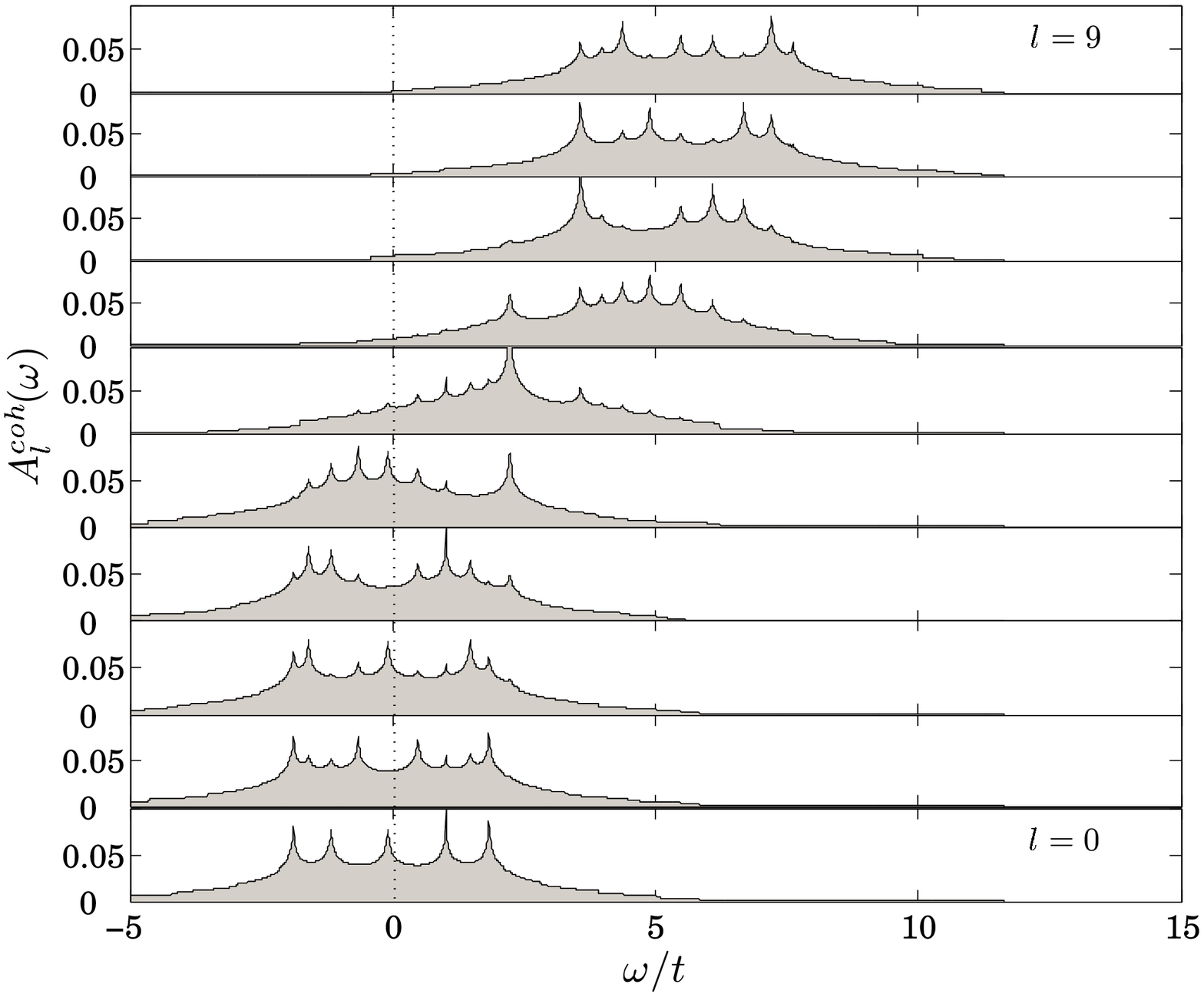}
\caption{Layer resolved spectral density for $U_{\mathrm{r}}=0$ for the $N=10$ heterostructure showed for $l=0,\dots, 9$.}
\label{fig:AcohU0}
\end{figure}
\begin{figure}
\centering
\includegraphics[width=1\linewidth]{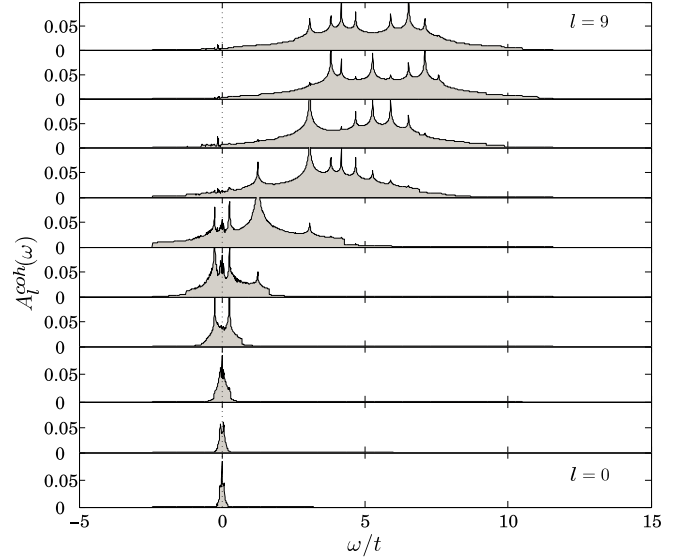}
\caption{Layer resolved spectral density for $U_{\mathrm{r}}=24t$ for the $N=10$ heterostructure showed for $l=0,\dots, 9$.}
\label{fig:AcohU24}
\end{figure}
Figure~\ref{fig:AcohU0} and \ref{fig:AcohU24} show the layer resolved spectral density $A_{l}^{\mathrm{coh}}(\omega)$ in the minimal interacting case and for $U_{\mathrm{r}}=24t$, respectively, for the $N=10$ heterostructure. In the latter case, the coherent part is strongly reduced towards the center of the heterostructure. The missing spectral weight  indicates nearly insulating behavior and that most of the spectral weight is transferred to upper and lower Hubbard band at a higher energy scale. (Note that the SBMF approximation is not able to catch the incoherent spectral weight which defines upper and lower Hubbard band.) However, outside the heterostructure, the spectral weight of the near-Fermi-surface-states is only weakly suppressed and almost identical to the minimal interacting case. We find that in the band-insulating material the chemical potential ($\omega=0$) is, in all cases studied, pinned to the the bottom of the conduction band. The highest-lying electron states are therefore only weakly bound (see also Ref.~\onlinecite{okamoto:075101}). The various singularities showing up in the spectral density enter here because of the van-Hoove singularity in the free electron density of states $N(\varepsilon)$ in combination with the infinite life-time of the quasiparticles at $T=0$. At finite temperature we expect that these singularities are washed out due to finite-life-time effects $\sim T^2$ of the quasiparticles. 

\begin{figure}
\centering
\includegraphics[width=1\linewidth]{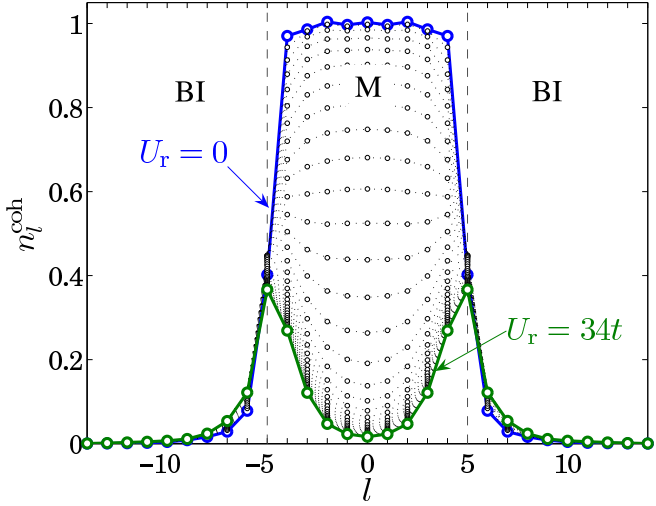}
\caption{(Color online) The coherent particle density $n_l^{\mathrm{coh}}$ of the $N=10$ heterostructure for various values of $U_{\mathrm{r}}=0,t,\dots,34t$.}
\label{fig:ncoh}
\end{figure}

By integration of Eq.~(\ref{eq:Alomega}) we find the coherent part of the particle density
\begin{equation}
n_l^{\mathrm{coh}}=\int_{-\infty}^{\infty}d\omega A_l^{\mathrm{coh}}(\omega)f_T(\omega)=z_l^2n_l
\end{equation} 
Figure~\ref{fig:ncoh} shows $n_l^{\mathrm{coh}}$ for various values of the on-site interaction $U_{\mathrm{r}}$.
It is interesting to note that the ratio $n_l^{\mathrm{coh}}/n_l=z_l^2=m_l^*/m$ gives the mass renormalization in layer $l$. A similar observation was mentioned in Ref.~\onlinecite{Okamoto:2005a} on the basis of quasiparticle dispersion calculations in the framework of DMFT. 
The coherent particle density clearly shows humps at the interfaces between Mott and band-insulating material whereas in the center of the heterostructure the coherent part is strongly suppressed for strong on-site interactions. These humps have a width of approximately $3a$ which suggests that the coherent quasiparticles responsible for the metallic behavior are confined to a very narrow region at the interface.
For large $N$ and $U_{\mathrm{r}}$ the coherent particle density is essentially independent of $N$ and $U_{\mathrm{r}}$.

\subsection{Drude weight and optical conductivity}
The zero frequency contribution to the optical conductivity at $T=0$ - the so called Drude weight - is one of the most relevant quantities both experimentally and theoretically to distinguish between an insulator and a metal. It was pointed out by Kohn\cite{Kohn:1964lr} that for a system with a (discrete) translation invariance in e.g.\  $x$-direction the real part of the complex optical conductivity $\sigma=\sigma'+i\sigma''$ has the form\cite{Millis:1990fk}
\begin{equation}
\sigma'(\omega)=D\delta(\omega)+\sigma_{\mathrm{reg}}'(\omega)
\label{eq:sigmagen}
\end{equation}
with $\lim_{\omega\rightarrow 0}\omega\sigma_{\mathrm{reg}}'(\omega)=0$. The weight of the Dirac delta function $\delta(\omega)$ is called the Drude weight $D$. $D$ provides an order parameter for the Mott transition at $T=0$ in non-disordered systems.\cite{Kohn:1964lr, Imada:1998qy} The system is metallic if $D>0$ and insulating if $D=0$. Studying the Drude weight therefore allows for an alternative assessment of the metallicity of the interface between band- and Mott- insulating materials.

As we have shown in detail in the previous sections the low-energy behavior of the system  is described, at least within the SBMF approximation, by coherent quasiparticles (see also the discussion in the last section of this paper, Sec.~\ref{sec:conclusions}). In this case, also the low-energy part of $\sigma(\omega)$ is expected to be dominated by the contributions of these quasiparticles.\cite{Baeriswyl:1987lr} A reasonable way to estimate the low-energy behavior of the optical conductivity is therefore to apply linear response theory to the renormalized mean-field Hamiltonian. In this way a close formal analogy to previous studies of the optical conductivity in the single-band Hubbard model can be drawn.\cite{Maldague:1977lr, Baeriswyl:1987lr, Shastry:1990lr, Wagner:1991lr, Scalapino:1992lr, Millis:1990fk} Alternatively, the Drude weight $D$ can be obtained from the ground-state energy of a system with twisted boundary conditions - an idea originally formulated by Kohn.\cite{Kohn:1964lr} In the Appendix \ref{ap:drude} we show that the same result for the Drude weight is directly obtained by considering twisted boundary conditions and using the SBMF approximation in order to estimate the dependence of the ground-state energy on the twist.

\begin{figure}
\centering
\includegraphics[width=0.8\linewidth]{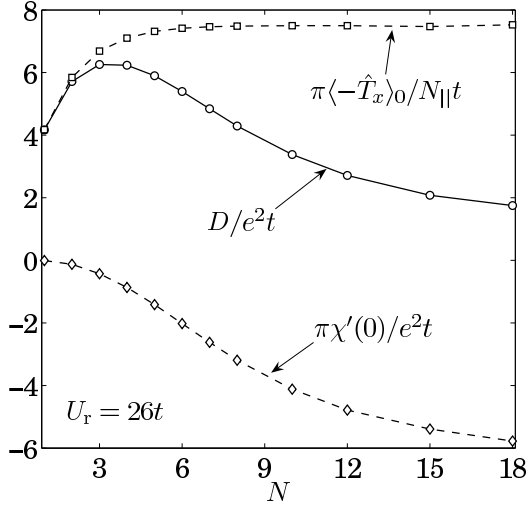}
\caption{The Drude weight [Eq.~(\ref{eq:dw})] and its two quasiparticle contributions as a function of $N$ for $U_{\mathrm{r}}=26t$.}
\label{fig:drudeN}
\end{figure}

In the following we choose units such that $\hbar=1=c$. The response of the system to a time dependent uniform electric field in the $x$-direction is found to be
\begin{equation}
\sigma(\omega)=\frac{i}{\omega+i0^+}\left[\frac{e^2}{N_{||}}\langle-\hat{T}_x \rangle_0
+\chi(\omega)\right].
\label{eq:sigma}
\end{equation}
The kinetic energy of the coherent quasiparticles
\begin{equation*}
\langle\hat{T}_x\rangle_0=\sum_{k\nu}\varepsilon_kn_{k\nu}^{\mathrm{coh}}
\end{equation*}
is given by the expectation value of the renormalized kinetic energy operator in $x$-direction
\begin{equation*}
\hat{T}_x=-2t\sum_{kl\sigma}\cos k_x z_l^2 \hat{f}_{lk\sigma}^{\dag}\hat{f}_{lk\sigma}^{\phantom{\dag}}.
\end{equation*}
The current-current correlation function is given by
\begin{eqnarray}
\chi(\omega)=\frac{1}{N_{||}}\sum_n\left|\langle n|\hat{J}_{\mathrm{p}x}|0\rangle\right|^2\left[\frac{1}{\omega-(E_n-E_0)+i0^+}\right.&&\nonumber\\
\left.-\frac{1}{\omega+(E_n-E_0)+i0^+}\right]&&
\label{eq:curcorr}
\end{eqnarray}
with the renormalized paramagnetic current in $x$-direction
\begin{equation*}
\hat{J}_{\mathrm{p}x}=e2t\sum_{kl\sigma}\sin k_x z_l^2\hat{f}_{lk\sigma}^{\dag}\hat{f}_{lk\sigma}^{\phantom\dag}.
\end{equation*}
The many-body states $\{| n\rangle\}$ in Eq.~(\ref{eq:curcorr}) are pseudo-fermionic Slater-determinants with energy $E_n$. We decompose $\chi$ into real and imaginary part, $\chi=\chi'+i\chi''$. The real part of $\sigma(\omega)$ is found to be of the form of Eq.~(\ref{eq:sigmagen}) with the Drude weight given by
\begin{equation}
D=\frac{\pi e^2}{N_{||}}\langle -\hat{T}_x \rangle_0+\pi\chi'(0)
\label{eq:dw}
\end{equation}
and a regular part $\sigma'_{\mathrm{reg}}(\omega)=-\chi''(\omega)/\omega$. 
The first term in Eq.~(\ref{eq:dw}) is positive and converges quite rapidly both with increasing $U_{\mathrm{r}}$ and $N$ because for large $N$ and $U_{\mathrm{r}}$ the coherent quasiparticles are located in a narrow region at the interfaces between the two materials, as discussed in the previous section. The rapid convergence is shown in Fig.~\ref{fig:drudeN} where the $N$-dependence of $\langle -\hat{T}_x \rangle_0$ for fixed $U_{\mathrm{r}}=26t$ is plotted.
The second term $\pi\chi'(0)$ in Eq.~(\ref{eq:dw}) is negative, thus it reduces the Drude weight. However, this weight is transferred to the regular part $\sigma_{\mathrm{reg}}$ at finite values of $\omega$. This is essentially a consequence of the $f$-sum rule for the real part of $\sigma(\omega)$
\begin{equation}
\int_0^{\infty}\sigma'(\omega)d\omega=\frac{\pi e^2}{2N_{||}}\langle -\hat{T}_x \rangle_0
\label{eq:frule}
\end{equation}
meaning that
\begin{equation*}
\frac{\pi\chi'(0)}{2}=\int_0^{\infty}\sigma'_{\mathrm{reg}}(\omega)d\omega.
\end{equation*}
In general $\chi(\omega)\not\equiv 0$ because it contains non-vanishing contributions from matrix elements  between band pairs of the same parity
\begin{equation*}
\langle \nu k|\hat{J}_{\mathrm{p}x}|\nu' k\rangle=e2t\sin k_x\sum_{\sigma l}z_l^2\psi_{k\nu}(l)\psi_{k\nu'}(l).
\end{equation*}
Hence, the fact that the renormalization factor $z_l$ depends on $l$ formally allows for interband transitions when a time-dependent electric field is applied in the $x$-direction. However, this is only possible for $U_{\mathrm{r}}>0$ - in contrast to the case of an electric field applied along the $z$-direction.\cite{okamoto:241104} As shown in more details in the Appendix \ref{ap:drude}, this is closely related to the fact that the renormalization of the quasiparticle dispersion $Z_{k\nu}$ [Eq.~(\ref{eq:Zknu})] is $k$-dependent. Therefore, unlike the case of the uniform Hubbard model, in systems with spatial inhomogeneity $\chi(\omega)$ acquires non-vanishing contributions from interband quasiparticle scattering processes when $U_{\mathrm{r}}>0$. For the real part at $\omega=0$ we explicitly find that
\begin{eqnarray}
\chi'(0)&=&-\frac{2}{N_{||}}\sum_{n>0}\frac{|\langle n|\hat{J}_{\mathrm{p}x}|0\rangle|^2}{E_{n}-E_{0}}\nonumber\\
&=&-\frac{2}{N_{||}}\sum_{k\nu'<\nu}\left|\langle \nu k|\hat{J}_{\mathrm{p}x}|\nu' k\rangle\right|^2\frac{f_0(E_{k\nu'})-f_0(E_{k\nu})}{E_{k\nu}-E_{k\nu'}}\nonumber\\
&=&\frac{e^2}{N_{||}}\sum_{k\nu}\frac{\partial Z_{k\nu}}{\partial\varepsilon_k}|\nabla\varepsilon_k|^2f_0(E_{k\nu}).
\end{eqnarray}
We expect that also $\chi'(0)$ saturates with increasing $U_{\mathrm{r}}$ and $N$ because in the large $N$ and strongly interacting limit essentially two independent interfaces contribute to the Drude weight $D$. However, as seen in Fig.~\ref{fig:drudeN}, the convergence of $\chi'(0)$ with increasing $N$ for fixed $U_{\mathrm{r}}$ is much slower than for $\langle-\hat{T}_x\rangle_0$.
\begin{figure}
\centering
\includegraphics[width=0.8\linewidth]{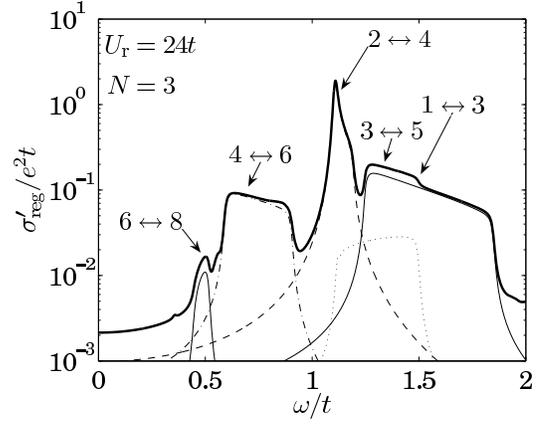}
\caption{Low-energy optical conductivity of the $N=3$ haterostructure in the strongly correlated case $U_{\mathrm{r}}=24t$. The main contributions to the different features can be assigned to transitions between two bands $\nu\leftrightarrow\nu'$ and are shown as thin lines.\emph{don't forget to add the prime...}}
\label{fig:sigmaregN3}
\end{figure}
The regular part of the low-energy optical conductivity $\sigma'_{\mathrm{reg}}(\omega)=-\chi''(\omega)/\omega$ can be written as
\begin{eqnarray}
\sigma_{\mathrm{reg}}'(\omega)=\frac{2\pi}{N_{||}}\sum_{k\nu'<\nu}\left|\langle \nu k|\hat{J}_{\mathrm{p}x}|\nu' k\rangle\right|^2&&\nonumber\\
\times[f_0(E_{k\nu'})-f_0(E_{k\nu})]\delta(\omega^2-(E_{k\nu}-E_{k\nu'})^2).&&
\label{eq:sigmareg}
\end{eqnarray}
Figure~\ref{fig:sigmaregN3} shows $\sigma_{\mathrm{reg}}'(\omega)$ of the $N=3$ heterostructure in the strongly interacting case $U_{\mathrm{r}}=24t$ as calculated from Eq.~(\ref{eq:sigmareg}) by approximating the $\delta$-function by a Lorentzian of width $0.01t$. The main contributions to the different features can be assigned to transitions involving only two different bands $\nu\leftrightarrow\nu'$ of the same parity. The width of the individual features is due to the $k$-dependent splitting of the bands and not due to the approximation of the $\delta$-function.

\begin{figure}
\centering
\includegraphics[width=0.8\linewidth]{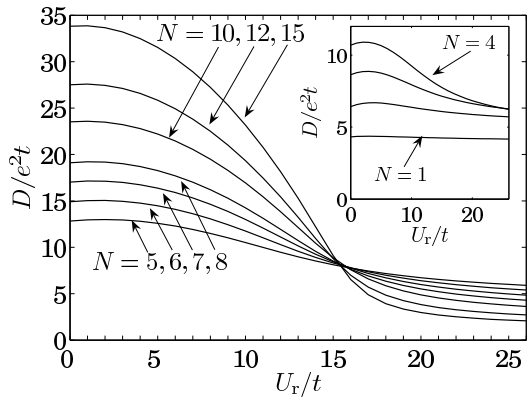}
\caption{Drude weight for different values of $N$ as function of $U_{\mathrm{r}}$.}
\label{fig:drudeU}
\end{figure}

The $U_{\mathrm{r}}$-dependence of $D$ is shown in Fig.~\ref{fig:drudeU} for different heterostructures of width $N$. Again we can distinguish between the Hartree regime and the Mott regime that are clearly separated by a common intersection point of $D(N,U_{\mathrm{r}})$ slightly below $U_{\mathrm{r}}\approx U_{\mathrm{c}}^{\mathrm{BR}}$ for $N\geq 5$. This two regimes are characterized by the different behavior of $D$ in the limit $N\rightarrow\infty$: In the Hartree regime, $D$ increases with increasing $N$ because the number of mobile carriers is proportional to $N$. In the Mott regime the $N$ dependence is much weaker and the main contribution to the Drude weight stems from the interface regions. $D$ decreases with growing $N$ and we expect that it saturates eventually.

\begin{figure}
\centering
\includegraphics[width=0.7\linewidth]{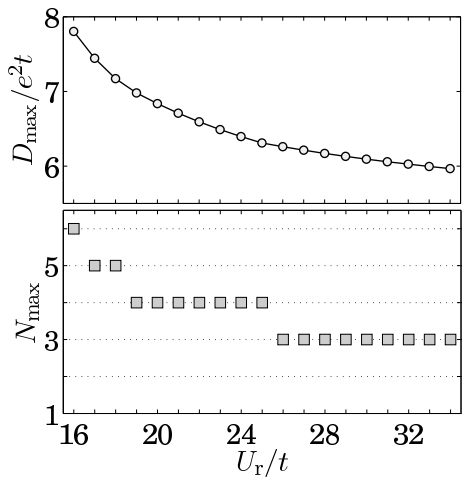}
\caption{The maximal value of the Drude weight and the corresponding $N$.}
\label{fig:drudemax}
\end{figure}

Because the dependence of $D$ on $U_{\mathrm{r}}$ is rather weak if $N$ is small ($1\sim 4$) there is another interesting aspect. Namely, in the Mott regime there is a finite value of $N$ which maximizes the Drude weight for a given interaction strength. We show in Fig.~\ref{fig:drudemax} $D_{\mathrm{max}}$ and $N_{\mathrm{max}}$ as function of $U_{\mathrm{r}}$. Both are decreasing functions of  $U_{\mathrm{r}}$.

\section{Conclusions}
\label{sec:conclusions}
To summarize, we have presented a mean-field study of a model heterostructure characterized by a BI/M/BI stacking (see Fig.~\ref{fig:model}) based on the Kotliar-Ruckenstein slave-boson theory. Although the model discussed here is too simple to make quantitative statements about the experimentally realized LaTiO$_3$/SrTiO$_3$ interface it allows to discuss some aspects of short-range correlation in spatially nonuniform systems. In particular, it suggests that the mobile carriers are confined to a relatively narrow region at the interface\cite{okamoto:241104} forming a quasi-two-dimensional electron gas as experimentally observed in LaTiO$_3$/SrTiO$_3$ heterostructures.\cite{Ohtomo:2002fk} 
 
The SBMF approximation offers a relative simple and transparent way to include both the self-consistent determination of the charge distribution as well as spatially varying hopping renormalization factors due to strong correlations. The basis of this approach is an effective one-dimensional Schr\"odinger equation in combination with self-consistency equations for the charge distribution and the double occupancy. Due to the lack of spatial homogeneity the quasiparticles obtain a nontrivial momentum dependence which can vary substantially by changing the on-site interaction $U$ or the width $N$ of the sandwiched material. This dependence was studied by calculating the renormalization of the quasiparticle dispersion and the quasiparticle weight.  Dynamical properties were studied by calculating  the layer resolved spectral density and the low-energy part of the optical conductivity including the Drude weight.
To which extend the quasiparticle properties discussed in this paper are reliable depends on the interesting question whether or not the quasi-two-dimensional electron gas formed at the interfaces is a Fermi liquid (see also the discussion in Ref.~\onlinecite{lee:075106}). Whereas Sr$_{1-x}$La$_x$TiO$_3$ shows clear Fermi liquid behavior with enhanced electron-electron scattering processes as $x\rightarrow 1$ (Ref.~\onlinecite{Tokura:1993lr}) and resistivity measurements on LaTiO$_3$/SrTiO$_3$ superlattices are consistent with this picture\cite{Shibuya:2004} it is an open problem if the Fermi liquid description survives when reducing the dimensionality further to a quasi-two-dimensional system. However, in the SBMF analysis of the model, the system \emph{is} a Fermi liquid with well defined quasiparticles and the present approach offers a rout to access to Fermi liquid parameters on the basis of a microscopic model. On the other hand, we confirmed various results reported in previous dynamical-mean-field studies\cite{okamoto:241104} which leads to the conclusion that the SBMF approximation gives a reasonable description of the low-energy part and is thus able to catch the most important aspects of correlations in spatially nonuniform situations. However, only on-site correlations can be taken into account and the introduction of a layer dependent mass renormalization distinctly increases the computational effort as compared to Hartree mean-field calculations. But for all that, compared to DMFT calculations, the SBMF approximation has the advantage that it is numerically less involved.

So far we did not treat other important aspects inherent in strongly correlated electron systems. The most conspicuous is the ignorance of magnetic ordering which especially in bipartite lattices is of relevance. To some extent this can be included also on the mean-field level by considering an enlarged unit cell. On the other hand, the results shown in this paper are basically independent of the details of the  bandstructure and may be suited to cases where magnetic order is suppressed - like on a triangular lattice. Nevertheless, inter-site correlations, namely spin-spin interaction, can not be treated within the present approach. This is of particular interest in connection with superconductivity arising in doped Mott insulators. A study of superconductivity is presented elsewhere. 

\begin{acknowledgments}
We would like to thank I.\ Milat, S.\ D.\ Huber, M.\ Indergand, T.\ M.\ Rice, A.\ Koga, A.\ Ferraz, A.\ J.\ Millis and S.\ Okamoto for many helpful discussions during different stages of this work. This study was financially supported by the Toyota Central R \& D Laboratories, Nagakute, Japan and the NCCR MaNEP of the Swiss Nationalfonds. 
\end{acknowledgments}

\begin{appendix}
\section{Avoided level crossing in two minimal models}
\label{ap:minmod}
If we consider a finite system in the $z$-direction composed of only $L=3$ or $L=4$ layers (but keeping the symmetry $z\leftrightarrow -z$) the effective Schr\"odinger equation (\ref{eq:Heff}) can easily be solved analytically.

\underline{$L=3$:}
The central layer may be almost localized with a strongly renormalized dispersion $\epsilon_0(\varepsilon)=z_0^2\varepsilon+\lambda_0$ where $z_0^2\ll1$. On the other hand, the two outer layers may be only weakly renormalized $\epsilon_{-1}(\varepsilon)=z_1^2\varepsilon+\lambda_1=\epsilon_1(\varepsilon)$ with $z_1^2\approx1$. The hybridization (interlayer hopping) is given by $\tilde{V}=-tz_0z_1$. This leads to the following bands
\begin{eqnarray}
E_1&=&(\epsilon_0+\epsilon_1)/2-\sqrt{(\epsilon_1-\epsilon_0)^2/4+2\tilde{V}^2},\nonumber\\
E_2&=&\epsilon_1,\nonumber\\
E_3&=&(\epsilon_0+\epsilon_1)/2+\sqrt{(\epsilon_1-\epsilon_0)^2/4+2\tilde{V}^2}\nonumber.
\end{eqnarray}
The renormalization of $E_1$ and $E_3$ is strongly $\varepsilon$-dependent. Interestingly, one band is unchanged by the finite hybridization. The corresponding eigenfunction $\psi_2=(1,0,-1)/\sqrt{2}$ has no weight in the center. Furthermore, in the $\varepsilon$-range where $E_{1,3}\approx E_2$, $\psi_{1,3}$ has also only little weight in the center.  The type of avoided level crossing in this minimal model corresponds to the type observed in heterostructures with even $N$ (see Fig.~\ref{fig:dispNALC}).

\underline{$L=4$:}
In this case the two central layers are characterized by $\epsilon_{-1/2}(\varepsilon)=z^2_{1/2}\varepsilon+\lambda_{1/2}=\epsilon_{1/2}(\varepsilon)$ and the outer layers by $\epsilon_{-3/2}(\varepsilon)=z_{3/2}^2\varepsilon+\lambda_{3/2}=\epsilon_{3/2}(\varepsilon)$. The interlayer hopping matrix elements are given by $V_0=-tz_{1/2}^2$ and $V_1=-tz_{1/2}z_{3/2}$ which leads to the following bands:
\begin{eqnarray}
E_1&=&\bar\epsilon+\frac{V_0}{2}-\sqrt{\left(\frac{\Delta\epsilon+V_0}{4}\right)^2+V_1^2},\nonumber\\
E_2&=&\bar\epsilon-\frac{V_0}{2}-\sqrt{\left(\frac{\Delta\epsilon-V_0}{4}\right)^2+V_1^2},\nonumber\\
E_3&=&\bar\epsilon+\frac{V_0}{2}+\sqrt{\left(\frac{\Delta\epsilon+V_0}{4}\right)^2+V_1^2},\nonumber\\
E_4&=&\bar\epsilon-\frac{V_0}{2}-\sqrt{\left(\frac{\Delta\epsilon-V_0}{4}\right)^2+V_1^2}\nonumber.
\end{eqnarray}
Here we use $\bar{\epsilon}=(\epsilon_{1/2}+\epsilon_{3/2})/2$ and $\Delta\epsilon=\epsilon_{1/2}-\epsilon_{3/2}$. The $\varepsilon$-dependence of these bands is shown in Fig.~\ref{fig:dispNALC}. The type of avoided level crossing is generic for heterostructures with odd $N$.
\section{Drude weight and twisted boundary conditions}
\label{ap:drude}
The Drude weight can be obtained from the ground-state energy of a system with twisted  boundary conditions. The idea originally formulated by Kohn\cite{Kohn:1964lr} is to consider periodic boundary conditions in e.g.\ $x$-direction and thread the system with a flux $\Phi$. The flux can be  represented by a vector potential $\mathbf{A}=(\hbar c/e) \Phi\mathbf{\hat{x}}/N_x$ (lattice constant $a=1$) where $N_x$ is the number of lattice sites in $x$-direction. We choose units such that $\hbar=c=1$. Then, due to the Peierls phase, the hopping in $x$-direction obtains an additional phase factor $t\rightarrow t\exp(\pm i\Phi/N_x)$
which changes the free dispersion
\begin{equation}
\varepsilon_k(\Phi)=-2t\left[\cos(k_x+\Phi/N_x) +\cos(k_y)\right].
\end{equation}
The Drude weight can be calculated from the ground-state energy density $E_{\mathrm{G}}(\Phi)$ as \cite{Kohn:1964lr,Shastry:1990lr,Millis:1990fk}
\begin{equation}
D=\pi e^2N_x^2\left.\frac{d^2 E_{\mathrm{G}}(\Phi)}{d\Phi^2}\right|_{\Phi=0}.
\end{equation}
We note here that the limit $\Phi\rightarrow0$ has to be taken before going to the thermodynamic limit $N_x\rightarrow\infty$. In order to estimate the dependence of $E_{\mathrm{G}}$ on the twist $\Phi$ we use the SBMF approximation. Because the explicit dependence of the saddle-point equations [Eq.~(\ref{eq:selfnP}, \ref{eq:selfdP}, \ref{eq:selflamP})] on $\Phi$ is in second order (the ground state does not carry any current) we find that the linear change of the mean-fields in $\Phi$ vanish
\begin{equation}
\left.\frac{d\, d_l}{d\Phi}\right|_{\Phi=0}=\left.\frac{d\, n_l}{d\Phi}\right|_{\Phi=0}=\left.\frac{d\,\lambda_l}{d\Phi}\right|_{\Phi=0}=0.
\end{equation}
The calculation of $D$ is thus considerably simplified because only the explicit dependence on $\Phi$ enters
\begin{eqnarray}
D&=&\pi e^2N_x^2\left.\frac{\partial^2 E_{\mathrm{G}}(\Phi)}{\partial\Phi^2}\right|_{\Phi=0}\nonumber\\
&=&-\frac{\pi e^2}{N_{||}}\sum_{k\nu}\left[\varepsilon_k n_{k\nu}^{\mathrm{coh}}-\frac{\partial Z_{k\nu}}{\partial\varepsilon_k}|\nabla\varepsilon_k|^2f_0(E_{k\nu})\right].
\label{eq:drudekohn}
\label{eq:drude}
\end{eqnarray}
where $n_{k\nu}^{\mathrm{coh}}$ is given in Eq.~(\ref{eq:nkcoh}) and $Z_{k\nu}$ in Eq.~(\ref{eq:Zknu}). This expression is equivalent to Eq.~(\ref{eq:dw}). Furthermore, in second order perturbation theory we find that
\begin{equation}
\frac{\partial Z_{k\nu}}{\partial \varepsilon_k}\equiv\frac{\partial^2 E_{k\nu}}{\partial\varepsilon_k^2}
=2\sum_{\nu'\neq\nu}\frac{\left[\sum_lz_l^2\psi_{k\nu}(l)\psi_{k\nu'}(l)\right]^2}{E_{k\nu}-E_{k\nu'}}
\end{equation}
which shows that the nonuniform renormalization of the quasiparticle dispersion is directly connected to the spatial variation of $z_l$.

The integration necessary for the second term in Eq.~(\ref{eq:drudekohn}) is most conveniently done by introducing the weighted density of states\cite{Denteneer:1995a}
\begin{equation}
N_{\mathrm{v}}(\varepsilon)=\frac{1}{N_{||}}\sum_{k}|\nabla\varepsilon_k|^2\delta(\varepsilon-\varepsilon_k)
\end{equation}
which for a square lattice can be expressed in terms of complete elliptic integrals of the first and secon kind
\begin{equation*}
N_{\mathrm{v}}(\varepsilon)=\frac{8t}{\pi^2}\left[E\left(1-\left(\frac{\varepsilon}{4t}\right)^2\right)-\left(\frac{\varepsilon}{4t}\right)^2K\left(1-\left(\frac{\varepsilon}{4t}\right)^2\right)\right]
\end{equation*}
for $|\varepsilon|\leq 4t$ and zero otherwise.
\end{appendix}
\bibliography{bib/references}
\end{document}